\documentclass[aps,prd,twocolumn,groupedaddress,showpacs]{revtex4}
\usepackage{graphics}
\usepackage{color}
\usepackage{ifthen}
\usepackage{amssymb}
\usepackage{tabularx}

\def\({\left(}
\def\){\right)}
\def\[{\left[}
\def\]{\right]}
\def\be{\begin{equation}}
\def\ee{\end{equation}}
\def\beq{\begin{eqnarray}}
\def\eeq{\end{eqnarray}}
\begin{document}
\title{Constraint on noncommutative spacetime from PLANCK data}

\author{Joby P.~K.}
\email{joby@iiap.res.in}
 
\author{Pravabati Chingangbam}
\email{prava@iiap.res.in}

\author{Subinoy Das}
\email{subinoy@iiap.res.in}

\affiliation{Indian Institute of Astrophysics, Koramangala II Block,
  Bangalore 560 034, India}

\begin{abstract}
We constrain the energy scale of noncommutativity of spacetime using CMB data from PLANCK. We find that PLANCK data put the lower bound on the noncommutativity energy scale to about 20 TeV, which is about a factor of 2 larger than a previous constraint that was obtained using data from WMAP, ACBAR and CBI. We further show that inclusion of data of $E$ mode of CMB polarization will not significantly change the constraint. 
\end{abstract}

\pacs{98.80.-k,98.80.Cq}

\maketitle
\section{Introduction} 

Physics at the Planck scale must incorporate quantized gravity and the correct way to describe physical events at that scale must be either a quantum theory of  gravity or string theory. These theories suggest that spacetime may be  noncommutative at a length scale of the order of Planck length. 

There is strong evidence from observational data that    inflation~\cite{Starobinsky:1979,Guth:1981,Starobinsky:1982,Linde:1982,Albrecht:1982}, a period of accelerated expansion in the very early history of the Universe, happened when the energy scale of the Universe was about three orders of magnitude lower than the Planck scale~\cite{Planck:2013kta}. 
Because of the accelerated expansion very small distances got stretched to cosmological sizes during inflation. If spacetime is noncommutative at very small length scales then such scales would have been stretched to possibly detectable cosmological sizes at the end of inflation. So, inflation provides a window of opportunity for probing Planck scale physics. Inflation is usually modeled to arise due to the evolution of a scalar field, the so-called inflaton field.  Quantum fluctuations of the inflaton field provide the initial seeds of gravitational potential wells around which matter could cluster and form structures. If spacetime is noncommutative the statistical properties of these quantum fluctuations must be affected by it. 

Fluctuations seen in the temperature and polarization of the cosmic microwave  background radiation~\cite{Penzias:1965,Cobe:1992} encode information about the physical properties of the Universe, its evolution and the properties of the primordial fluctuations generated during inflation. 
The CMB is possibly the cleanest means to understand the properties of the primordial fluctuations. Investigations on the effect of spacetime noncommutativity on inflation and signatures in the CMB have been carried out in several  works~\cite{Chu:2001,Lizzi:2002,Brandenberger:2002,Huang:2003,Tsujikawa:2003,Barosi:2008,Fatollahi:2006-1,Fatollahi:2006-2,Fabi:2008ta,Akofor:2008,Akofor:2009,Karwan:2009ic,Moumni:2009dc,Shtanov:2009wp,Malekolkalami:2010bc,Koivisto:2010fk,Soda:2012zm,Nautiyal:2013bwa}. In~\cite{Akofor:2009} the authors obtained a constraint on the noncommutativity parameter using data from  WMAP~\cite{WMAP5:1,WMAP5:2,WMAP5:3}, ACBAR~\cite{ACBAR:1,ACBAR:2,ACBAR:3} and CBI~\cite{CBI:1,CBI:2,CBI:3,CBI:4,CBI:5}. 

In this paper we investigate the effect of noncommutativity of spacetime on the two-point correlations of the CMB temperature fluctuations along the same line as~\cite{Akofor:2009}. 
We first analyze the effect of the noncommutativity parameter on the theoretical CMB angular power spectrum. We find that its effect is most pronounced in the range of scales probed by the PLANCK mission and hence, PLANCK data are best suited for constraining it. 
We further analyze the effect on the angular power spectrum of the $E$ mode of CMB polarization. Then we perform Monte Carlo Markov chain (MCMC) analysis to obtain the constraint on the 
noncommutativity parameter using data from PLANCK~\cite{Planck:2013kta},  baryon acoustic oscillations (BAO) from SDSS-DR9~\cite{Padmanabhan:2012hf,Anderson:2012sa} and the 6dFGS~\cite{Beutler:2011hx} and BICEP2~\cite{Ade:2014xna}. 
We obtain about a factor of 2 tighter constraint on the noncommutativity parameter in comparison to the value obatined in~\cite{Akofor:2009}. The improved constraint comes primarily from the higher resolution and accuracy of PLANCK compared to WMAP, ACBAR and CBI.

This paper is organized as follows.  
In Sec. II we briefly review the basics of noncommutative spacetime, quantum fields on such a spacetime and the effect on inflaton fluctuations. In Sec. III we discuss the consequence on the angular power spectrum of temperature fluctuations of the CMB. In Sec. IV we discuss the MCMC analysis and present our results. We end with a summary and discussion in section V. 

\section{Noncommutative spacetime and its effect on CMB angular power spectra}

In this section we briefly review the basic equations of noncommutative spacetime and how its effect manifests in the angular power spectrum of the CMB temperature fluctuations and $E$ mode of polarization. We follow the formulation in~\cite{Akofor:2008,Akofor:2009}. If spacetime is noncommutative the coordinate operators,  $\hat x^{\mu}$, do not commute and instead take the form
\begin{equation}
[{\hat x}^{\mu}, {\hat x}^{\nu}] = i\theta^{\mu\nu},
\label{eqn:xcom}
\end{equation}
where $\theta^{\mu\nu}$ is an antisymmetric constant matrix with dimension $L^{2}$. Let $\vec\theta^0\equiv \left( \theta^{01},\theta^{02},\theta^{03}\right)$ be in the third direction. Then $\vec\theta^0 =\theta \hat\theta^0 $, where $\hat\theta^0$ is the unit vector $(0,0,1)$. 
This relation is not invariant under Lorentz transformation (but is invariant under a deformed Lorentz symmetry). For cosmological application 
a natural choice of coordinates is the one where the time coordinate measures the proper time for an observer in a galaxy and the spatial coordinates measure the physical distances in such spatial slices. So what enters on the right-hand side of Eq.~(\ref{eqn:xcom}) is the physical  $\theta^{\mu\nu}$ which is constant throughout the history of the Universe. 

If instead we choose conformal coordinates, Eq.~(\ref{eqn:xcom}) will have the same form, with $\hat x^{\mu}$ now being coordinates in the conformal frame and $\theta^{\mu\nu}$ will no longer be time independent.  $\theta^{\mu\nu}$  in the two frames will be related to each other as 
\begin{eqnarray}
\theta^{0i,{\rm ph}} &=& a(t) \theta^{0i,{\rm co}} (t), \nonumber\\
\theta^{ij,{\rm ph}} &=& a^2(t) \theta^{ij,{\rm co}} (t),
\label{eqn:thetaph}
\end{eqnarray}
where $a(t)$ is the scale factor of expansion of space and the upper scripts are ``ph'' for physical and `co' for conformal. 
We carry out the computations here in the comoving frame and only in the end translate the result into the physical frame. For simplicity of notation we use $\theta^{\mu\nu}$ where we actually mean $\theta^{\mu\nu,{\rm co}} (t)$.  

In noncommutative spacetime the requirement of consistent statistics leads to  deformation of the quantum fields, $\phi_{\theta}$, in comparison to fields on commutative spacetime, $\phi_0$.
When applied to the context of cosmological scalar perturbations generated during inflation the deformation of the mode expansion of the fields leads to a primordial power spectrum of scalar perturbations, whose form is modified from the corresponding expression in commutative spacetime, given by
\begin{equation} 
P_{\Phi_{\theta}}({\bf k}) = P_{\Phi_{0}}(k) \cosh\left( H\vec\theta\,{}^0\cdot{\bf k}\right),
\label{eqn:PPhitheta}
\end{equation}
where $P_{\Phi_{0}}(k)$ is the usual direction independent power spectrum in the commutative case given by
\begin{equation} 
P_{\Phi_0}(k) = \frac{A_s}{k^{3}} \left(\frac{k}{k_0} \right)^{n_s-1},
\end{equation}
with $A_s$ being the amplitude, $n_s$ the spectral index and $k_0$ a pivot scale that can be suitably chosen. Then, the angular power spectrum of the CMB temperature fluctuations is given by 
\begin{equation}
C^{TT}_{\ell} = \int {\rm d}k\, k^2 P_{\Phi_0}(k) |\Delta^T_{\ell}(k)|^2 i_0(\theta H k ),
\label{eqn:cl_TT}
\end{equation}
where the function $i_0$ is the modified spherical Bessel function of the first kind, given by $i_0(x) = \sinh(x)/x$, $H$ is the Hubble parameter during inflation. For details of the calculations of Eqs.~(\ref{eqn:PPhitheta}) and (\ref{eqn:cl_TT}) we refer to~\cite{Akofor:2008,Akofor:2009}. 
$\Delta^T_{\ell}(k)$ is the transfer function for temperature fluctuations which encodes the physical processes relevant within the causal patches of the Universe during the decoupling and subsequent epochs. In writing this equation we have implicitly assumed that $\Delta_{\ell}(k)$ is not affected by the noncommutavity of spacetime and it is reasonable to do so since the noncommutativity scale is expected to be much smaller than the scales at which the transfer function is relevant.

A small fraction of the CMB photons, those that had their last scattering from free electrons towards the end of the decoupling epoch, are polarized due to velocity gradients in the plasma. The polarization degrees of freedom are usually decomposed into the so-called $E$ and $B$ modes~\cite{Kamionkowski:1996ks,Zaldarriaga:1996xe}. Fluctuations of the $E$ mode originate from the primordial scalar perturbations while fluctuations of the $B$ mode arise from primordial tensor perturbations. Actually $E$ mode fluctuations also depend on primordial tensor perturbations but the effect is subdominant since the amplitude of tensor fluctuations is much smaller than that of scalar fluctuations. Hence we ignore it here. The effect of noncommutativity on the $E$ mode of polarization can then be deduced along the same lines as done for temperature fluctuations in~\cite{Akofor:2009}. The angular power spectrum of $E$ mode is obtained to be 
\begin{equation}
C^{EE}_{\ell} = \int {\rm d}k\, k^2 P_{\Phi_0}(k) |\Delta^{E}_{\ell}(k)|^2 i_0(\theta H k ),
\label{eqn:cl_EE}
\end{equation}
where $\Delta^E_{\ell}(k)$ is the corresponding transfer function function for $E$ mode.
\section{Effect of $\theta$ on the CMB angular power spectra}

In this section we estimate the effect of $\theta$ on $C^i_{\ell}$, where the index $i$ refers to either `$TT$' or `$EE$'. To show the results we use the variable $D^i_{\ell}$ which is related to $C^i_{\ell}$ as  
\begin{equation}
D^i_{\ell}\equiv\frac{\ell (\ell +1)}{2\pi} C^i_{\ell}. 
\label{eqn:dl}
\end{equation}
We denote  $\alpha\equiv \theta H$ and study how varying $\alpha$ affects $D^i_{\ell}$. For calculating $C^i_{\ell}$ we use the publicly available cosmological Boltzman code \texttt{CAMB}~\cite{Lewis:2002ah,cambsite}. \texttt{CAMB}  essentially calculates the respective transfer functions, $\Delta^i_{\ell}(k)$, for temperature fluctuations and for the $E$ and $B$ modes of polarization. Then it integrates over $k$ to get the $C_{\ell}'s$, using Eq.~(\ref{eqn:cl_TT}), (\ref{eqn:cl_EE}) and a corresponding equation for the $B$ mode, for commutative spacetime without the $i_0$ factor. For our purpose, we modify  \texttt{CAMB} to incorporate $i_0$  in the expression for the primordial  scalar power spectrum.  
We calculate $D^i_{\ell}(\alpha)$ for different values of $\alpha$ and then calculate the difference $\Delta D^i_{\ell}(\alpha) \equiv D^i_{\ell}(\alpha) - D^i_{\ell}(\alpha=0)$ to see how the difference changes systematically as we vary $\alpha$.  
\begin{figure}[h]
\begin{center}
\resizebox{3.in}{2.6in}{\includegraphics{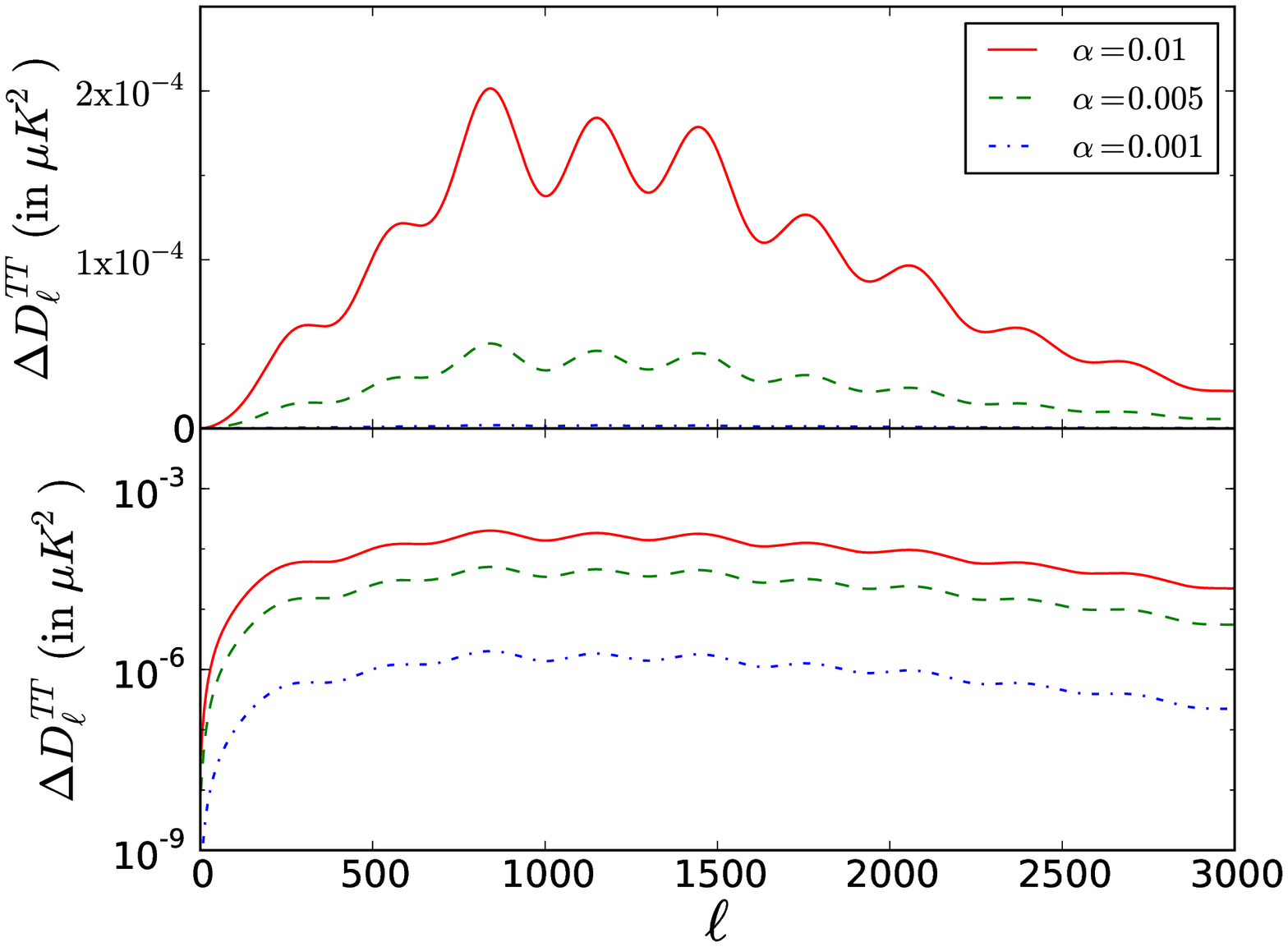}}
\resizebox{3.in}{2.6in}{\includegraphics{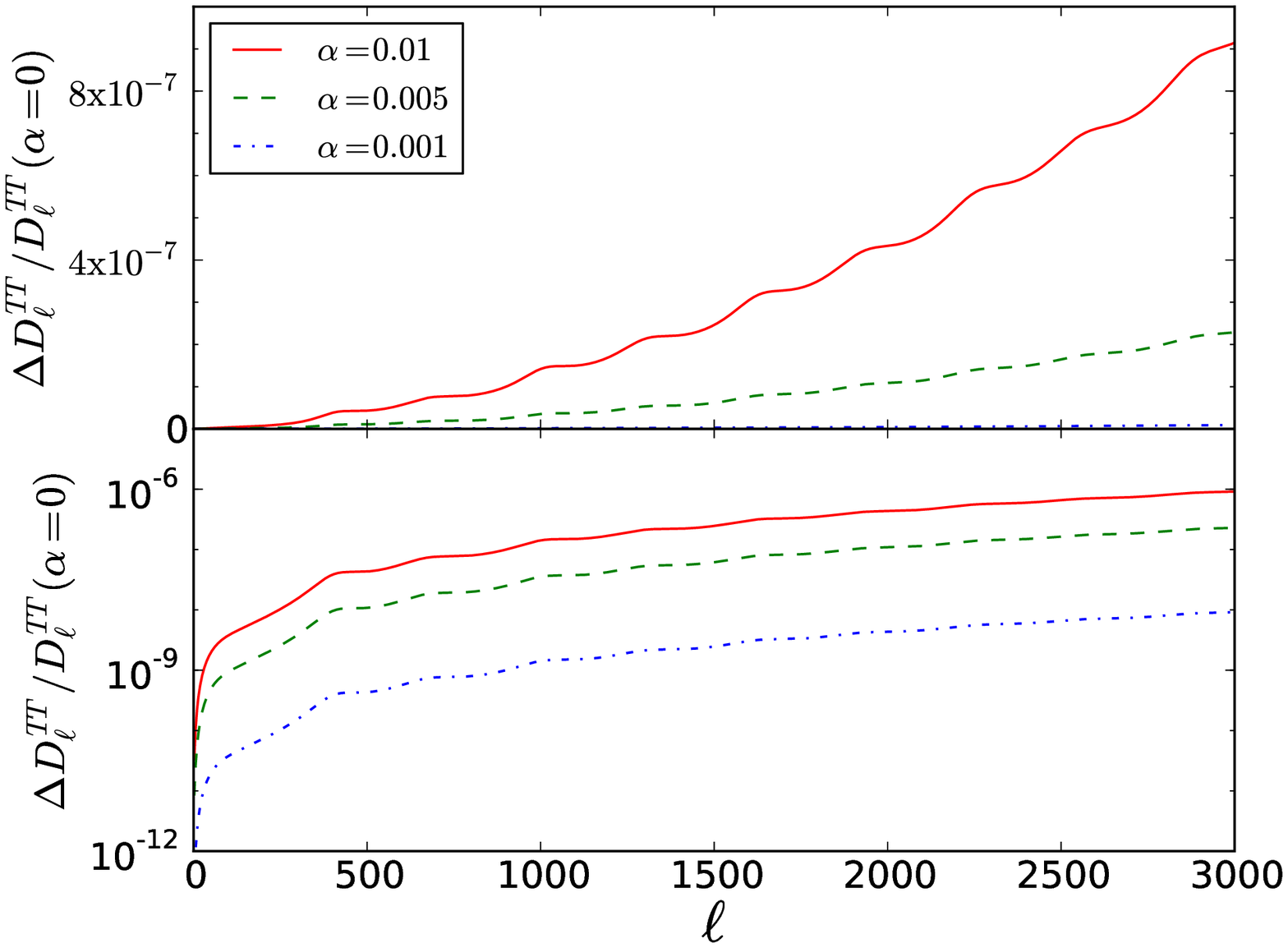}}
\end{center}
\caption{The plots show how $\Delta D^{TT}_{\ell}$ varies with $\alpha\equiv \theta H$. The unit for $\alpha$ is Mpc. {\em Top panel}: $\Delta D^{TT}_{\ell}\equiv D^{TT}_{\ell}(\alpha\ne 0) - D^{TT}_{\ell}(\alpha=0)$ with linear and log scale on $y$-axis. {\em Bottom panel}: $\Delta D^{TT}_{\ell}/D^{TT}_{\ell}(\alpha=0) $ with linear and log scale on $y$ axis.}
\label{fig:DCl_TT}
\end{figure}
\begin{figure}[h]
\begin{center}
\resizebox{3.in}{2.6in}{\includegraphics{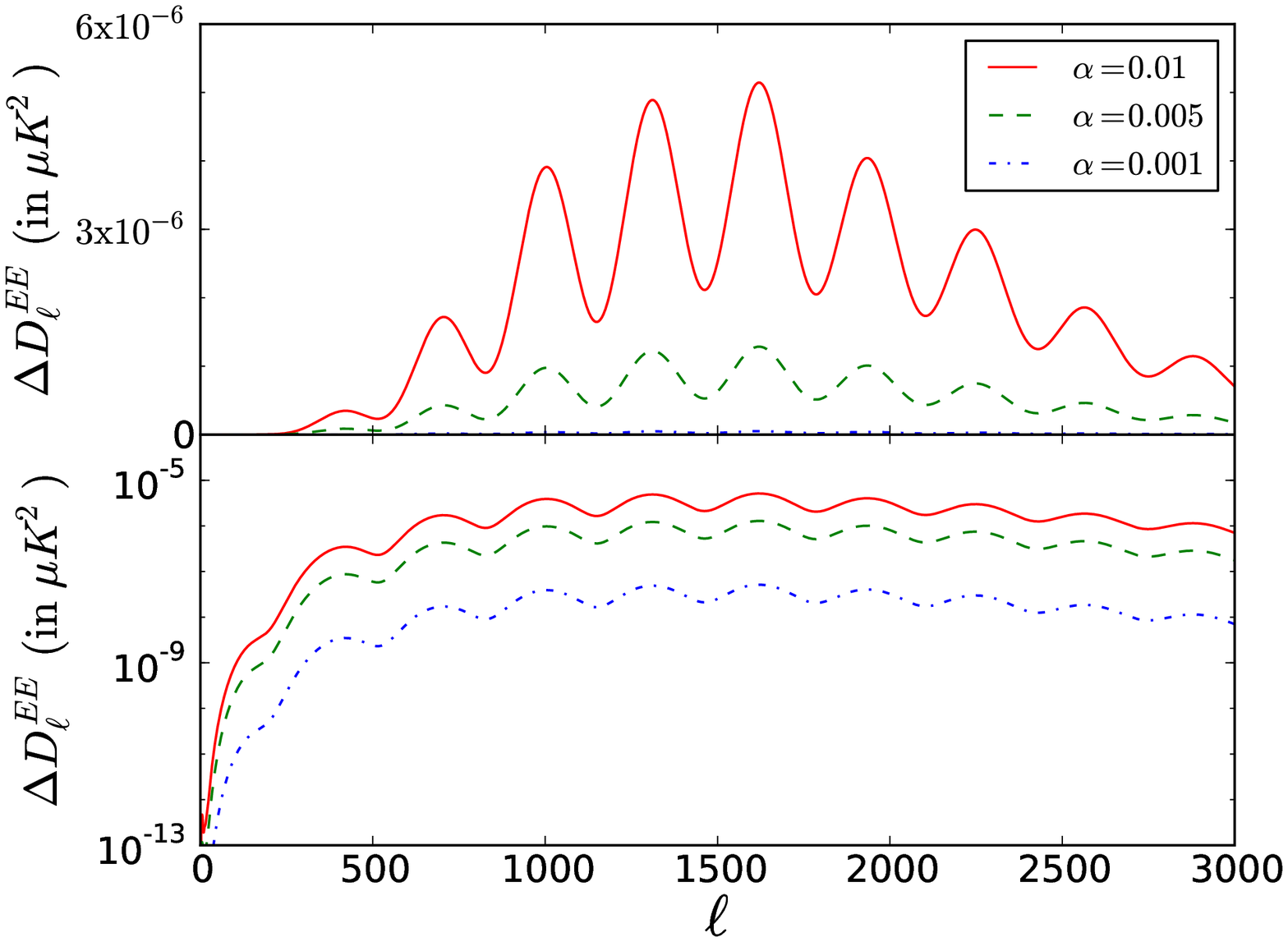}}
\resizebox{3.in}{2.6in}{\includegraphics{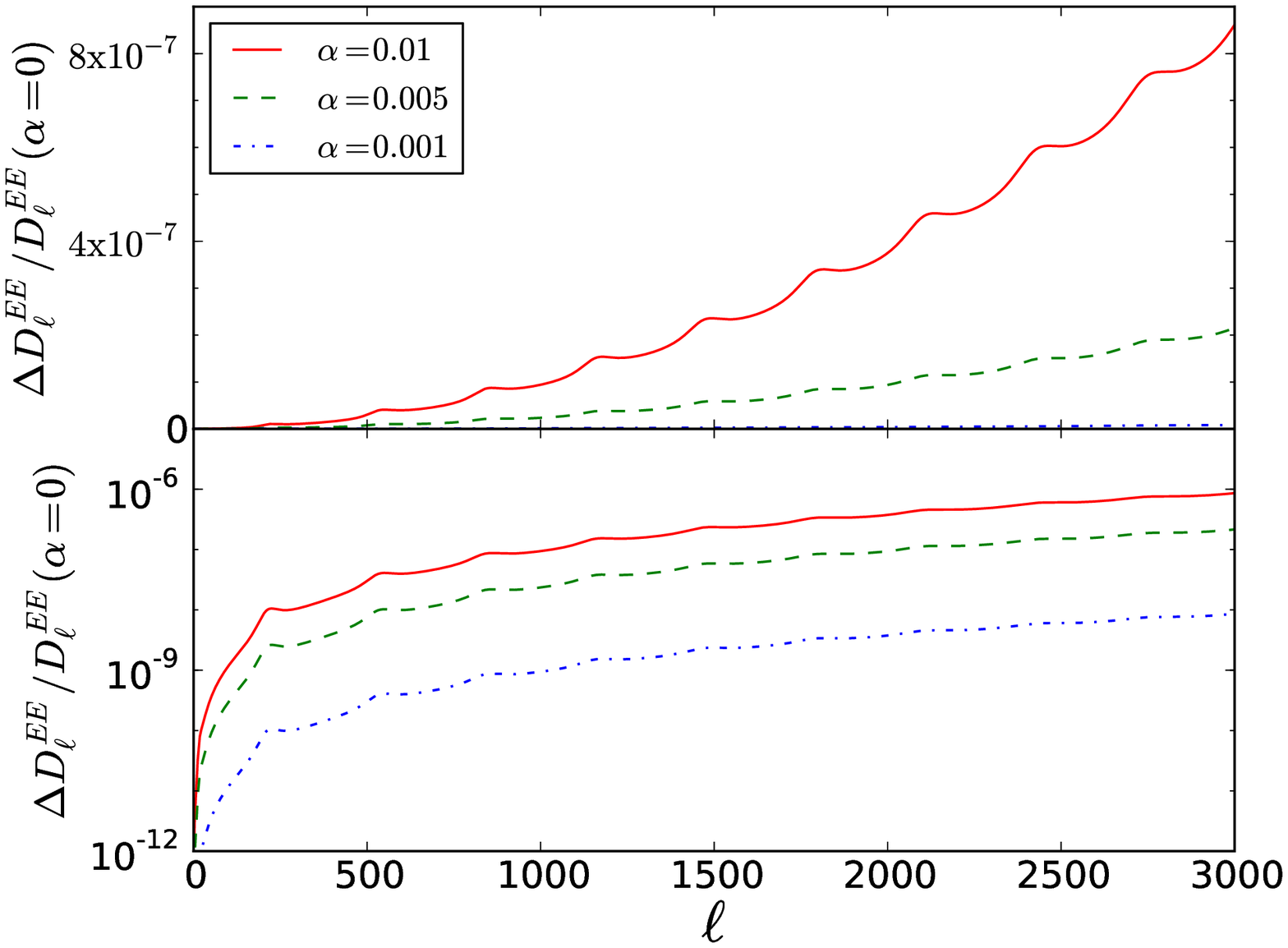}}
\end{center}
\caption{The plots show how $\Delta D^{EE}_{\ell}$ varies with $\alpha$. 
 Top panel: $\Delta D^{EE}_{\ell}\equiv D^{EE}_{\ell}(\alpha\ne 0) - D^{EE}_{\ell}(\alpha=0)$ with linear and log scale on $y$-axis. Bottom panel: $\Delta D^{EE}_{\ell}/D^{EE}_{\ell}(\alpha=0) $ with linear and log scale on $y$ axis.}
\label{fig:DCl_EE}
\end{figure}

In Figure~\ref{fig:DCl_TT} we have plotted $\Delta D^{TT}_{\ell}(\alpha)$ for different values of $\alpha$. \texttt{CAMB} uses units where $k$ is given in ${\rm Mpc}^{-1}$ and so $\alpha$ is in units of Mpc. The $\Lambda$CDM cosmological parameters used are the PLANCK best-fit values~\cite{Planck:2013kta}.  
Since $i_0(x)$ asymptotically goes to one as $x \rightarrow 0$ and monotonously increases as $x$ increases,  we observe an overall increase in the amplitude of $C_{\ell}$ as $\alpha$ increases. As seen in the top panel of Figure~(\ref{fig:DCl_TT}) we find that $\Delta D^{TT}_{\ell}(\alpha)$ is largest in the multipole range $<3000$, which is the range probed by PLANCK. The deviation amplitude drops sharply as $\alpha$ decreases and in the top lower panel of the same figure we have shown  $\Delta D^{TT}_{\ell}(\alpha)$ with log scale on the $y$-axis to highlight the sharpness of the amplitude drop. The plots in the bottom panel are $\Delta D^{TT}_{\ell}(\alpha)/D_{\ell}(\alpha=0)$ with linear and log scale on the $y$ axis and they bring out the effect of $i_0$. 
In~\cite{Akofor:2008}, five years data from WMAP which give information up to about $\ell_{\rm max}\sim 839$ were used to constrain $\theta$. Our calculation here shows that we can expect significantly tighter  constraint if we use data from PLANCK.  

Figure~(\ref{fig:DCl_EE}) shows the effect of $\alpha$ on $D^{EE}_{\ell}$ and all the panels are similar to those in Figure~(\ref{fig:DCl_TT}). We find that the amplitude of $\Delta D^{EE}_{\ell}$ is just over two orders of magnitude lower than seen for temperature fluctuations. The ratio  $D^{EE}_{\ell}/D^{EE}_{\ell}(\alpha=0)$ has similar values. From this we conclude that though polarization data provide independent information about the properties of the Universe we cannot expect significant improvement on constraints on $\theta$ from $E$ mode data. 

\section{Constraint on $\theta$ from data from PLANCK, BAO and BICEP2}

Based on the above discussion, in this section we compare the standard  $\Lambda$CDM cosmological model with noncommutative spacetime with data of CMB temperature fluctuations from PLANCK. We also use BAO data from SDSS~\cite{Padmanabhan:2012hf,Anderson:2012sa} and 2DFGS~\cite{Beutler:2011hx}, and CMB polarization data from BICEP2~\cite{Ade:2014xna}.
The cosmological parameters are the baryon density $\Omega_{b} h^2$, cold dark matter density $\Omega_{c} h^2 $, the Hubble parameter $H_0$, optical depth to reionization $\tau$, the amplitude $A_s$ and spectral index $n_s$ of primordial scalar fluctuations. In addition, we include the noncommutative parameter $\alpha \equiv \theta H $. 

We perform a Bayesian statistical inference analysis using the publicly available MCMC cosmological parameter estimation package \texttt{COSMOMC}~\cite{Lewis:2002ah,cosmomcsite}. To calculate the theoretical angular power spectra for temperature fluctuations and $E$ mode, which go in as inputs for the likelihood analysis, we use our modified version of \texttt{CAMB} described in the previous section. We adopt the Gelman-Rubin convergence criterion $R-1<0.02$ when generating multiple Markov chains, where $R$ is the {\em variance of chain means} divided by the {\em mean of chain variances}.  
We use the default flat priors given in \texttt{COSMOMC} for the standard model parameters . For $\alpha$ we use a flat prior with the upper limit value $0.1$ and lower limit $10^{-8}$. The upper limit was chosen so that we have a range wide enough to accommodate the constraint obtained by~\cite{Akofor:2009}. For the lower limit we did several runs of \texttt{COSMOMC} with progressively smaller values starting from $10^{-4}$ and ensured that our results are stable.   We run \texttt{COSMOMC} using various combinations of observational data sets. Here we present results for (a) PLANCK data alone, (b) PLANCK+BAO data, (c) PLANCK+BAO+BICEP2 data. The largest multipole for PLANCK data is $\ell_{\rm max} = 2479$.

\subsection{best-fit values and joint constraints} 

\begin{figure}[h]
\begin{center}
\resizebox{3.in}{3.7in}{\includegraphics{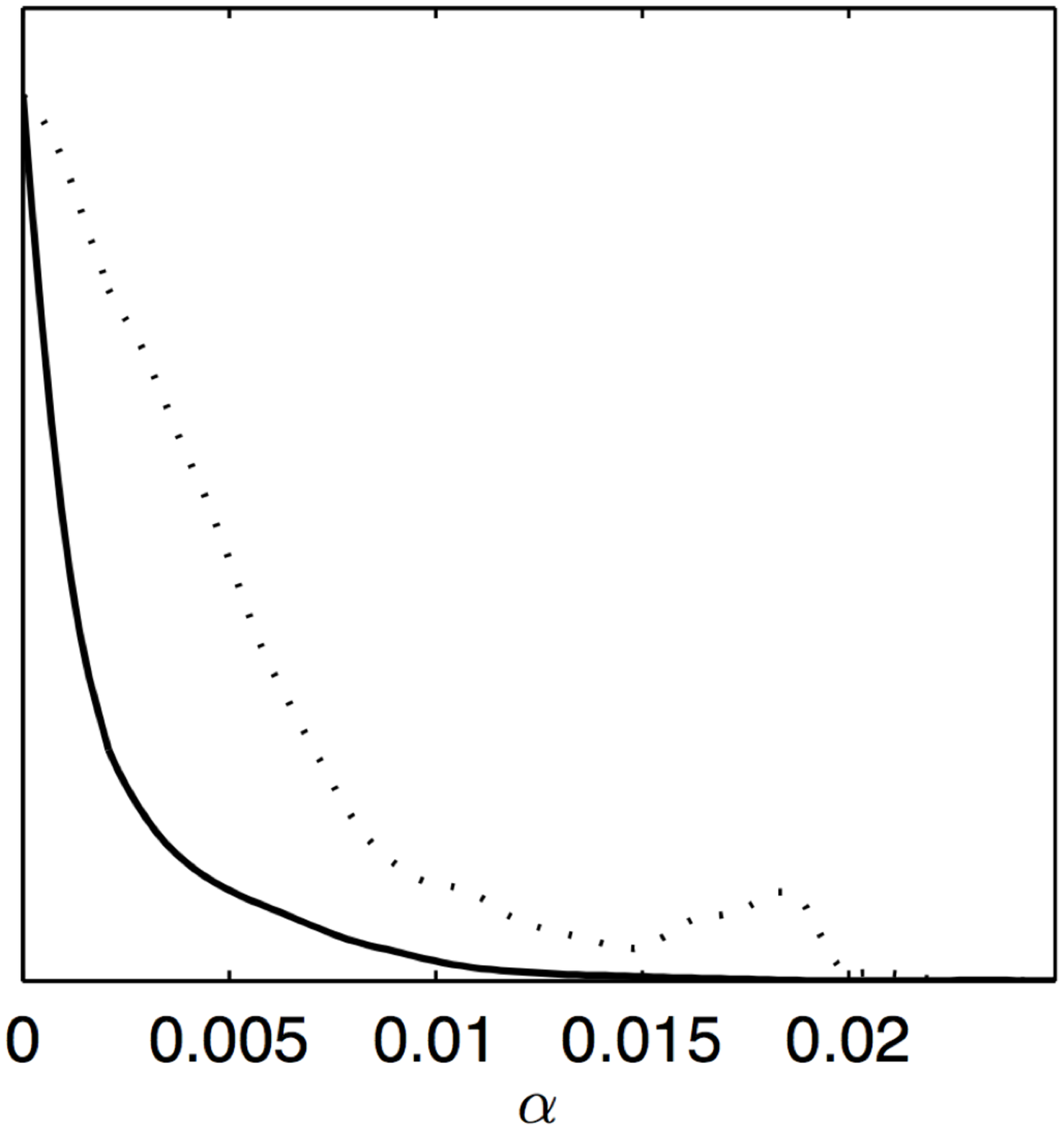}}
\end{center}
\caption{ The one-dimensional mean (the dashed lines) and marginalized (solid lines)
likelihood curves for $\alpha$, obtained using PLANCK data alone.}
\label{fig:1d}
\end{figure}

We first describe the result obtained using PLANCK data alone. In Figure~\ref{fig:1d} we have plotted the one-dimensional mean (dashed lines) and marginalized (solid lines) likelihood curves for $\alpha$. 
We observe that $\alpha$  is only constrained to take values below an upper bound.  
For a comparison we show the best-fit values and 2-$\sigma$ confidence limits of the $\Lambda$CDM parameters obtained in the noncommutative case with those from the commutative case, along with the $1$- and 2-$\sigma$ upper bounds for $\alpha$, in Table~I.
We find that the parameters change only marginally and, hence, they are not affected by the noncommutativity of spacetime. We get the  limits on $\alpha$ to be $\alpha  < 0.0026$ and $\alpha < 0.0087$ at  1-$\sigma$ and  2-$\sigma$, respectively. 

\begin{table}[t]
\begin{center}
\begin{tabular}{|l|c|p{3cm}|}\hline
Parameter & PLANCK   & PLANCK\\ 
{} & Commutative spacetime &  Noncommutative spacetime \\ \hline
$\Omega_{\rm b} h^2$ & $0.0220^{\,\, 0.0226}_{ \,\,0.0215}$ &   
$ 0.0219^{0.0225}_{0.0214}  $ \\ \hline

$\Omega_{\rm c} h^2 $ 	& $0.1199^{0.1251}_{ 0.1147} $ & $ 0.1198^{0.1252}_{0.1147}$ \\ \hline

$H_0$  & $67.29^{69.65}_{64.98}$ &   
$67.23^{69.53}_{64.90}$          \\  \hline

$\tau$  & $0.0898^{0.1158}_{0.0640} $&   
$0.0878^{0.1134}_{0.0622}$ \\ \hline

$n_s$  &  $0.9605^{0.9748}_{0.9462} $   &    
$ 0.9599^{0.9738}_{0.9457}$   \\ \hline

${\rm{ln}}(10^{10} A_s)$	& $ 3.090^{3.139}_{3.043}  $  & $3.084^{3.134}_{3.036}$ \\ \hline

$\alpha\equiv \theta  H $  &  \textemdash & 	
$ < 0.0026 $ (1-$\sigma$)  \\ 
{} &   \textemdash & $< 0.0087$ (2-$\sigma$) \\
\hline 
\end{tabular}
\vspace{0.3cm}
\caption{The mean values and the 2-$\sigma$ confidence level limits for the $\Lambda$CDM parameters in commutative spacetime and in noncommutative spacetime, obtained using PLANCK data alone.  
For $\alpha$ we show the 1-$\sigma$ and 2-$\sigma$ upper constraints. 
The pivot scale used for $A_s$ is $k=0.05\ \textrm{Mpc}^{-1}$.}
\end{center}
\label{tab:planck}
\end{table}

\begin{figure}[h]
\begin{center}
\resizebox{2.3in}{4.5in}{\includegraphics{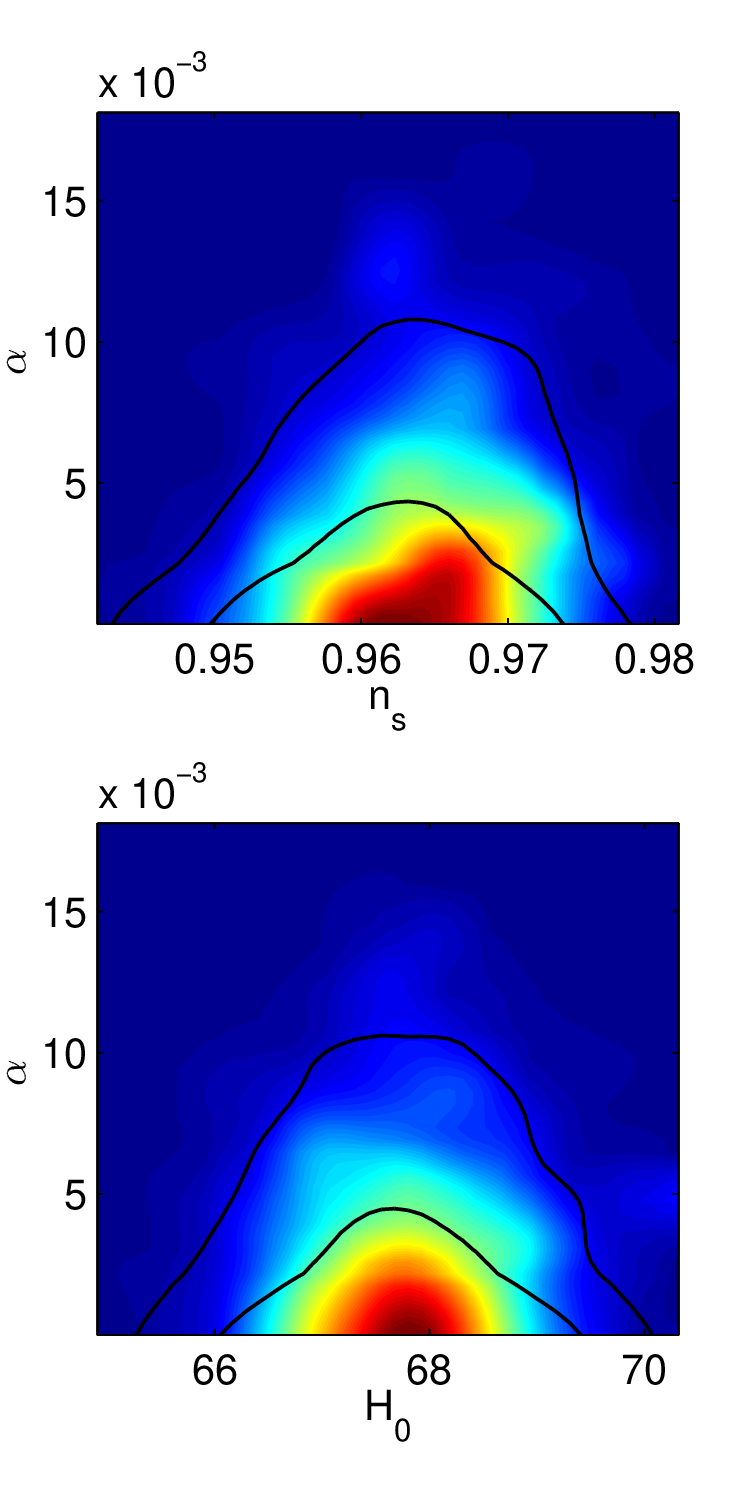}}
\end{center}
\caption{The 1-$\sigma$ and 2-$\sigma$ two-dimensional joint constraints on $\alpha$, $n_s$ and $H_0$, obtained using PLANCK data alone.} 
\label{fig:2d}
\end{figure}

Since $i_0$ is a monotonously increasing function one may expect degeneracy between $\alpha$ and $n_s$, and to a very mild extent with $H_0$. To investigate this we have plotted the 1- and 2-$\sigma$ two dimensional joint constraints between these three parameters in Figure~\ref{fig:2d}. We find only a very mild degeneracy between them and the strength of the degeneracy is larger for larger values $\alpha$, as can be seen from the 2-$\sigma$ contour. 

We repeat the analysis for noncommutative spacetime using PLANCK+BAO and PLANCK+BAO+BICEP2 data sets. The mean values and the 2-$\sigma$ confidence level limits for the cosmological parameters and 1- and 2-$\sigma$ upper bounds for $\alpha$ are shown in Table~\ref{tab:planck+}. There are only marginal shifts in the 1- and 2-$\sigma$ constraints on $\alpha$ after including these data sets. Inclusion of BICEP2 data slightly tightens the constraints on $\alpha$.

\begin{table}[t]
\begin{center}
\begin{tabular}{|l|c|p{3cm}|}\hline
Parameter & PLANCK + BAO \,\,\,\,\,\, &  PLANCK + BAO + BICEP2 \\ \hline 
$\Omega_{\rm b} h^2$ & $0.0220^{\,\, 0.0225}_{ \,\,0.0215}$ &   $ 0.0219 ^{ 0.0224}_{0.0214}  $ \\ \hline
$\Omega_{\rm cdm} h^2 $ 	& $0.1187^{0.1223}_{ 0.1153} $ & $ 0.1196^{0.1231}_{0.1162}$ \\ \hline
$H_0$  & $67.71 ^{66.10}_{69.25}$&   $67.30^{68.82}_{65.81}$          \\  \hline
$\tau$  & $0.0896^{0.1151}_{0.0657} $&   $0.0964^{0.1233}_{0.0717}$ \\ \hline
$n_s$  &  $0.9622^{0.9733}_{0.9504} $   &     $ 0.9600_{0.9487}^{0.9709}$   \\ \hline
${\rm{ln}}(10^{10} A_s)$	& $ 3.08^{3.13}_{3.03}  $  & $3.10^{3.15}_{3.05}$ \\ \hline
$\alpha\equiv \theta  H $  & $ < 0.0025$ (1-$\sigma$)   & 	$ < 0.0021$ (1-$\sigma$)  \\ 
{} & $< 0.0083$ (2-$\sigma$) & $< 0.0074$ (2-$\sigma$ \\
\hline 
\end{tabular}
\vspace{0.3cm}
\caption{The mean values and the 2-$\sigma$ confidence level limits for the cosmological parameters in noncommutative spacetime, obtained using PLANCK+BAO and PLANCK+BAO+BICEP2 data sets. For $\alpha$ we show the 1-$\sigma$ and 2-$\sigma$ upper constraints. 
The pivot scale used for $A_s$ is again $k=0.05\ \textrm{Mpc}^{-1}$.}
\label{tab:planck+}
\end{center}
\end{table}

\subsection{Constraint on the physical noncommutativity parameter}

We now convert the constraint on $\alpha$ to the corresponding constraint on $\theta$. We use the results for PLANCK data alone. We need to first fix the value of $H$ during inflation. Using the relation $P_T=H^2/2\pi$, where $P_T$ is the power spectrum of primordial tensor perturbations, and the tensor-to-scalar ratio $r\equiv P_T/P_{\Phi}$ we have
\begin{equation}
H^2= (2\pi )^2 r P_{\Phi}.
\label{eqn:Hubble}
\end{equation}
Using the best-fit value of $A_s$ from the third column of Table~I, and the 2-$\sigma$ limit $r<0.11$ from PLANCK~\cite{Planck:2013kta} in Eq.~(\ref{eqn:Hubble}) we obtain $H$ to be
\begin{equation}
H<1.943 \times 10^{-5} M_P,
\label{eqn:Hlimit}
\end{equation}
 where $M_P$ is the Planck mass. From the 1-$\sigma$ limit on $\alpha$ given in the second column of Table (I) we obtain the  1-$\sigma$ limit on $\theta$ to be $\theta < 0.427\times 10^{-9} m^2$. Using the 2-$\sigma$ limit on $\alpha$, we get  2-$\sigma$ limit on $\theta$ to be $\theta < 1.406\times 10^{-9} m^2$.
 
In order to obtain $\theta^{\rm ph}$, we use the first line of Eq.~(\ref{eqn:thetaph}). For this we need the value of the cosmological scale factor, $a$, at the time when inflation ended. 
As done in~\cite{Akofor:2009}, if we assume that the temperature of reheating of the Universe was close to the GUT energy scale, $10^{16}$GeV, then the scale factor at the end of inflation is roughly $a\simeq 10^{-29}$. Using this we obtain the constraint on the noncommutativity parameter to be 
$\sqrt{\theta^{\rm ph}}< 0.653 \times 10^{-19} m$ at 1-$\sigma$ and 
$\sqrt{\theta^{\rm ph}}< 1.186 \times 10^{-19} m$ at 2-$\sigma$. These  correspond to lower bounds on the noncommutativity energy scale given by  18.9764 and 10.4548 TeV at 1- and 2-$\sigma$, respectively. Note that these numbers will scale correspondingly if the reheating temperature is different. Comparing our 1-$\sigma$ bound with that in~\cite{Akofor:2009} we find a factor of 2 improvement in the constraint on the noncommutativity length scale. 

\section{Conclusion}

We confront the prediction of noncommutative spacetime, which is essentially captured by a modification of the power spectrum of primordial scalar perturbations, with observational data from PLANCK, BAO and BICEP2. We first analyze the effect of spacetime noncommutativity on the theoretical angular power spectrum of temperature fluctuations and $E$ mode of polarization of the CMB. We find that the effect is most pronounced in the multipole range $<3000$, implying that PLANCK data will provide more stringent constraint in  comparison to using WMAP data. We also find that the effect on the angular power spectrum of $E$ mode is much weaker than in temperature fluctuations and hence inclusion of $E$ mode data will not significantly improve the constraint.     

To obtain the constraint on the spacetime noncommutativity scale, $\theta^{\rm ph}$, we compare the theoretical CMB angular power spectrum of temperature fluctuations with the observed one obtained from PLANCK data. We perform a Bayesian analysis to obtain the best-fit values of the $\Lambda$CDM parameters and the spacetime noncommutativity parameter by performing MCMC likelihood analysis. We find that $\theta^{\rm ph}$ is constrained to be smaller than  $6.23 \times 10^{-20} \,m$ at 1-$\sigma$ confidence level which is a factor of 2 improvement from the previous limit obtained using data from WMAP, ACBAR and CBI. The improvement comes from the higher angular resolution of PLANCK in comparison to WMAP, and higher precision in comparison to ACBAR and CBI even though they probed comparable smaller angular scales.  

Since the effect of the spacetime noncommutativity decreases at multipoles higher than 3000 it is expected that including CMB data at smaller scales will not improve the constaint significantly. Further, our finding that the effect on the angular power spectrum of $E$ mode is much weaker than in temperature fluctuations suggests that inclusion of $E$ mode data will also not significantly improve the constraint. The effect of spacetime noncommutativity also affects the power spectrum of the primordial tensor modes~\cite{Shiraishi:2014owa} and will show up in the angular power spectrum of $B$ mode polarization. In our analysis we have not included this effect. This will not impact our results since the BICEP2 data are restricted to lower $\ell$ values where the effect of spscetime noncommutativity is negligible. 
The results of the BICEP2 experiment have been recently updated in a joint analysis by the BICEP2 and PLANCK teams~\cite{Joint_BICEP_PLANCK:2015} and there is no statistical significant detection of $B$ mode. This does not affect our result here.
In the future the inclusion of high-resolution data of $B$ modes can potentially tighten the constraint on $\theta^{\rm ph}$, subject to improvement in our ability to disentangle physical effects that are important at smaller scales such as gravitational lensing due to large-scale structure.  

\section*{Acknowledgments}
The computation required for this work was carried out on the Hydra cluster at IIA. We acknowledge use of the \texttt{CAMB} and \texttt{COSMOMC} packages. P.~C. would like to thank K.~P.~Yogendran for useful discussions. S.~D. would like to thank D.~Boriero and J.~Ramakers for useful suggestions regarding the \texttt{COSMOMC} run. 


\end{document}